\documentstyle[12pt]{article}
\input epsf
\begin{document}
 
\title{
Polarization asymmetry and the Higgs boson production mechanisms } 
 
\author{ \\ M.Dubinin \\
{\small \it Institute of Nuclear Physics, Moscow State University}\\
{\small \it 119899 Moscow, Russia}\\ \\
Y.Kurihara and Y.Shimizu\\
{\small \it National Laboratory for High Energy Physics (KEK)}\\ 
{\small \it Tsukuba, Ibaraki 305, Japan}}
\date{}
\maketitle
 
\begin{abstract}
We calculate the cross section of Higgs boson production 
in the reaction $e^+e^- \rightarrow
\nu \bar \nu b \bar b$ for the case of
left and right longitudinally polarized  electron
beam. Complete set of signal and irreducible background diagrams 
is considered. 
 
We propose a new critical test to identify the Higgs boson provided by  
the ratio of cross sections with left and right beam
polarization. This quantity is sensitive to the ratio of Higgs-$W$ and
Higgs-$Z$ couplings, since it 
exhibits the interplay of bremsstrahlung
and fusion mechanisms of Higgs boson production inherent to the Standard
Model scheme.  
\end{abstract} 
\newpage 
 
\section{Introduction}
 
Direct experimental observation of the Higgs boson would be
one of the most important evidences in favour of the 
Standard Model (SM) \cite{SM} scheme. For this reason much
experimental and theoretical effort has been applied to the problems
of Higgs boson signal observation and the calculation of
corresponding particle distributions.
 
The absence of Higgs signal in the $Z$ boson decays at LEP I  
sets the lower limit of the Higgs boson mass around 60 GeV
\cite{LEP100}. In the nearest future the possibilities to observe the Higgs
particle will be given by the experiments at LEP II and
Tevatron colliders. At LEP II it is expected to observe the light
Higgs boson ($M_H <$ 120 GeV) signal in the four fermion semi-leptonic final
states containing $b$-quark pair $l \bar l b \bar b$ \cite{LEPII}. 
Search for the signal in the four fermion states could be performed 
in the following at
next colliders \cite{ee500}. Such states
can be separated from the large hadronic backgrounds by the efficient
$b$-tagging procedure \cite{btag}.
 
Complete sets of tree level diagrams for the reactions
$e^+ e^- \rightarrow \nu_e \bar \nu_e b \bar b$, $e^+ e^- b \bar b$
contain two "signal" diagrams with intermediate Higgs boson line (see the
first row of diagrams in Fig.1) which are usually named as bremsstrahlung
\cite{Hbrems} and fusion \cite{fusion} mechanisms of Higgs production.
Besides signal graphs the complete tree level $2 \rightarrow 4$ body
$l \bar l b \bar b$   
amplitude contains many other irreducible background diagrams (24, 21
and 48 for $l=\mu, \quad \nu_e$ and $e$, correspondingly). Irreducible
background contribution is not small and in order to understand
completely the problem of signal separation from the bias it is not
sufficient to calculate only signal mechanisms. It is highly desirable
to consider bremsstrahlung and fusion mechanisms as the interfering parts
of the same amplitude summing all diagrams coherently. Such calculations
have been performed in 
\cite{CTLC1} ($\mu^+\mu^- b \bar b$ channel),
\cite{CTLC2} ($\nu \bar \nu b \bar b$ channel) and
\cite{CTLC3} ($e^+ e^- b \bar b$ channel) 
by means of eight-dimensional phase space Monte Carlo integration of the
complete tree level amplitude.
Semi-analytical results for the $\mu^+\mu^- b \bar b$ channel were obtained
in \cite{Bardin} by means of symbolical integration over
six angular variables and numerical integration over the
remaining two invariant masses. 
 
The calculations for $\nu \bar \nu b \bar b$ channel showed that it
has the largest cross section for the signal and a very interesting
interplay between the bremsstrahlung and fusion mechanisms \cite{CTLC2,BDD}.
At the energies of LEP II (175-205 GeV) and $\sqrt{s} \sim M_H+M_Z$
(threshold region) the contributions of two mechanisms to the cross
section are of the same order additionally enhanced by the positive
interference between them. At larger energy Higgs bremsstrahlung is the main 
one up to the
energy of several hundred GeV, when fusion begins to dominate. In the case of
$e^+ e^- b \bar b$ channel (when also two signal mechanisms contribute) 
the picture is less interesting. First,
the background from ladder diagrams (multipheripheral mechanisms) is
100 times greater than the signal \cite{CTLC3} and we need a special
procedure of signal separation. Second, the interference term between two
mechanisms is negative compensating fusion contribution
 in the threshold region. 
 
\section{ Higgs boson identification and polarization asymmetry}
 
In the situation of limited experimental statistics and many additional
factors affecting the separation and reconstruction of the signal (initial
state radiation, beamstrahlung, final state radiation, efficiency of
$b$-tagging, simulation of jet fragmentation, simulation of detector) it
is necessary to calculate several experimental observables giving the
possibility of unambiguous identification of the underlying mechanisms.
If some new resonance appears in the two $b$-jet invariant mass
distribution at Next Linear Collider, additional 
identification tests will be necessary to recognize
the SM Higgs particle in it.
If the signal from a new scalar particle with the main decay
channel to $b{\bar b}$ pair shows up in the future experiments, it will be 
not sufficient for identification of that particle as 
the Higgs boson. At the first stage comparison of the measured and 
calculated total 
production rates of the scalar particle can be used for identification. 
However, total rate has systematic errors from luminosity measurement, 
$b$-tagging efficiency and other sources. Moreover, in the nonminimal 
schemes it depends on
the number of Higgs doublets and the mixing angles of Higgs mass 
eigenstates. 
 
It is well known that polarized electron beams allow
us to reach higher precision in the measurement of SM parameters otherwise
unreachable \cite{pLEPI} and impose stronger limits on the deviations
from the SM scheme given by new physical phenomena. Extremely
interesting physical observable is the polarization asymmetry \cite{asymm}
 
\begin{eqnarray}
A_{LR}=\frac{\sigma(e_L^- e^+ )-\sigma(e_R^- e^+) }
            {\sigma(e_L^- e^+ )+\sigma(e_R^- e^+) }
\end{eqnarray}
where $e_R \, (e_L)$ denote right(left) longitudinal polarization
of the electron beam. This quantity is much less affected by
initial and final state radiative corrections than the total cross section 
and has no uncertainties
appearing from the hadronization of final quarks. These corrections are
cancelled in the ratio.
 
In the case of reaction $e^+ e^- \rightarrow \nu_e \bar \nu_e b \bar b$
left and right polarized electron beams distinguish the bremsstrahlung and
fusion mechanisms of Higgs boson production. In so far as the weak charged
current includes only left spinors the right polarization of initial
electron beam switches off the fusion mechanism. If we denote the
coupling constants of the Higgs boson to gauge bosons by ${\cal G}_{\small WWH}$
and ${\cal G}_{\small ZZH}$ and factor them out, we can write for two
signal diagrams
\begin{eqnarray}
\sigma(e_L^- e^+ )&=& {\cal G}^2_{\small WWH}F_W^L+ 
{\cal G}_{\small WWH}{\cal G}_{\small ZZH} F_{WZ}^{L}  
+{\cal G}^2_{\small ZZH}F_Z^L,\\
\sigma(e_R^- e^+ )&=& {\cal G}^2_{\small ZZH}F_Z^R.
\end{eqnarray}   
Fusion diagram has additional electroweak coupling $g$ in comparison
with brems\-stra\-hlung absorbed into $F$. 
Defining $A'_{LR}$ as the ratio of the right to the left cross section
we get 
\begin{eqnarray}
A'_{LR}&=&\frac{\sigma(e_R^- e^+) }
            {\sigma(e_L^- e^+ ) } \\
~&=& \left(R^2 \frac{F_W^L}{F_Z^R} + R \frac{F_{WZ}^{L}}{F_Z^R}
+\frac{F_Z^L}{F_Z^R} \right)^{-1}, 
\end{eqnarray}
where
\begin{eqnarray}
R=\frac{{\cal G}_{\small WWH}}{{\cal G}_{\small ZZH}}.
\end{eqnarray}
The interference term is negligible under some conditions
we shall discuss later.   
One can expect very precise measurement of $A'_{LR}$ at the energies
of Next Linear Colliders. Cancellation of systematical errors from
luminosity measurement and the effects of beamstrahlung in this
quantity seems especially attractive. On the other hand, $F_{L,R}$
can be accurately calculated and thus we can precisely define the factor 
$R$ which in the SM is equal to  
\begin{equation}
R=\cos^2 \vartheta_W. 
\end{equation}
This relation is essential for the Higgs boson identification.
In the SM (7) is a property of couplings of the Higgs-gauge field
interaction lagrangian with spontaneously broken symmetry. 
In the case of polarized beams, a measurement of $A'_{LR}$ directly
implies the verification of (7) on a better level of
precision than by using total production rate, where the systematics
and beam effects are not cancelled. 
 
Since the relation (7) is independent of the number of Higgs doublets, 
it is satisfied also for CP-even Higgs bosons $h$ and $H$ of the
MSSM. Additional factors in the $WWh(H)$ and $ZZh(H)$ vertices (mixing of
Higgs mass eigenstates) are cancelled in the asymmetry
for $h(H)$ boson production case and in this sense the dependence
of $A'_{LR}$ is universal for the interplay of two 
mechanisms. The possibility to distinguish SM from MSSM using the
polarised asymmetry measurement does not appear.

\section{Cross sections in the case of polarized beams and
         the polarization asymmetry}
 
Calculation of the amplitude has been performed simultaneously
by means of two computer packages - GRACE \cite{GRACE} and
CompHEP \cite{CompHEP}. GRACE uses massive helicity amplitude method
for numerical calculation of 23 diagrams (Fig.1), 
while CompHEP uses projection
operators for energy and helicity states to calculate symbolic expressions
for 23$\times$(23+1)/2=276 squared diagrams and interferences between them. 
Eight dimensional integration over the four particle 
phase space 
has been done with the help of adaptive Monte-Carlo
integration package BASES \cite{BASES}.
More details about both packages can be found in \cite{LEPII}.
 
The basic input parameters used in calculations are
\begin{eqnarray}
M_W &=& 80.20~{\rm GeV},  \nonumber \\
M_Z &=& 91.16~{\rm GeV}, \nonumber\\
\Gamma_Z &=& 2.53~{\rm GeV}, \nonumber\\
m_b &=& 5 ~{\rm GeV}, \nonumber\\
\cos^2 \vartheta_W &=& \frac{M_W^2}{M_Z^2}. \nonumber
\end{eqnarray}
The results are presented for three different Higgs boson masses 
with the corresponding tree-level widths   
\begin{eqnarray}
M_H &=& 110,~130,~150~{\rm GeV}, \nonumber\\
\Gamma_H &=& 6.12,~7.26,~8.40~{\rm MeV}. \nonumber
\end{eqnarray}
 
We used the "fixed width" (or "naive") prescription for the insertion
of exact (Breit-Wigner form) propagators in the complete tree level
amplitude. Generally speaking, the result can be strongly dependent on
the prescription \cite{BDD}, especially in the threshold energy region.
Latest discussion of the subject can be found in \cite{width}. Fixed
width method seems to be more adequate to the situation than other
methods. 
 
First we calculate the cross sections for two signal
diagrams only (first row in Fig.1).
Total cross sections for the individual signal diagrams in the processes   
$e^-_L e^+ \rightarrow \nu \bar \nu b \bar b$ and
$e^-_R e^+ \rightarrow \nu \bar \nu b \bar b$
are shown in Fig.2. Here three neutrino species 
($\nu=\nu_e,\nu_{\mu},\nu_{\tau}$) are summed up.
The fusion mechanism (dotted lines in Fig.2a)
is switched off  
in the $\nu_e \bar \nu_e b \bar b$
channel by right polarization of the electron beam. 
For muon and tau neutrino channels
only the Higgs bremsstrahlung exists.
 
However, this picture is unphysical because we have not taken into
account the interference between signal mechanisms and neglected the irreducible
background diagrams. Main background comes from $e^+ e^- \rightarrow
Z^* Z^*$, $Z^* \gamma$ mechanisms. In order to suppress their contribution 
we introduced   
the $M(b \bar b)$ invariant mass cut of $\pm 1$ GeV around the Higgs resonance. 
The result of complete tree level calculation is shown in Fig.3. 
Even at the threshold region, $\sigma(e^-_{\small L} e^+)$ is about
twice larger than the $\sigma(e^-_{\small R} e^+)$. Left polarisation
of electron beam enhances the total rate.
 
For the $e_{\small L}^- e^+ \rightarrow \nu_e {\bar \nu_e} b {\bar b}$ case 
at $M_H$=110 GeV the contributions from 
bremsstrahlung, fusion, and interference terms separately are 
shown in Fig.4.
We can see that even with the strong cut the signal interference 
and bias graphs
cannot be completely neglected.  
For instance, in the left polarized beam case at $\sqrt{s}=$ 220 GeV 
and $M_H=$ 110 GeV the bremsstrahlung 
gives 71 fb, fusion diagram squared gives 12 fb and about 3 fb comes
from signal interference and the irreducible background.  
At $\sqrt{s}=$ 340 GeV destructive interference reduces the
"pure signal" cross section by $-3.3$ fb. 
 
It is interesting to notice that in the case of polarized beams the
signal interference term $F_{WZ}$ in (5) changes sign and at some energy
rather far from the threshold $M_H+M_Z$ it is equal to zero. 
If the contribution of irreducible
background diagrams is small, the experimental measurement of $A'_{LR}$  
is directly related to the Higgs-gauge boson coupling ratio (6).
 
We show the correction to polarization asymmetry $A'_{LR}$ originating
from the signal interference and irreducible background diagrams 
in Fig.5. At the energies near
the threshold $M_H+M_Z$ the correction exceeds 40\%. However, near the  
zero point of signal-signal interference it is almost negligible indeed.

Polarization asymmetry $A'_{LR}(s)$ calculated with initial
state radiative corrections (ISR)
is shown in Fig.6. 
We are using the structure 
function approach in this   
calculation \cite{ISR}. The normalization and 
shape of $A'_{LR}$ energy
dependence are affected around 10\% by ISR. 
Fortunately there is a flat region of a CM energy dependence of the cross 
section as shown in Fig.3. In these regions the effect of the ISR is very small.
Moreover these regions coincide with zero-interference regions.
 
We have chosen a benchmark point at $M_H$=110 GeV 
and the CM energy of 230 GeV and
estimated an expected accuracy of the $A'_{\small LR}$ measurement.
At this point an effect of the interference term and initial state radiation
is very small.
We assume that 0.1\% accuracy can be obtained in the calculation of
functions $F$ 
in equation (5) by using our event generators and
the detailed detector simulation and 0.2\% accuracy 
can be reached for $\cos^2 \vartheta_W$ measurement.
A main source of the systematic error of the measurement may come from
uncertainty of the polarization measurement.
1\% accuracy was achieved in SLAC experiment at SLC\cite{pLEPI}. 
The expected accuracy of the
$A'_{\small LR}$ measurement with 1.0, 0.5, and 0.1\% errors of the 
polarization measurement are shown in Fig.7 as a function of the integrated
luminosity after the detector acceptance correction. 
The same values of the integrated luminosity for the left and right polarized
beams are assumed.
One can expect approximately
2\% accuracy of the polarized asymmetry measurement.

\section{Conclusion}
 
If some new signal similar to that expected from the Higgs boson will
be observed at Next Linear Collider, it would need to be studied in
more details. At LEP II energies the only observables of detailed study
that were discussed \cite{LEPII} are the angular distribution of the Higgs boson
from bremsstrahlung, which is expected to be rather flat, and similar to
this the angular distribution of Higgs decay products (reconstructed
in the Higgs decay frame).  
 
Electron beam polarization provides us a very useful tool to
study the mechanisms of Higgs boson production. The measurement of left-right
asymmetry (4) for the semi-leptonic four fermion channels $\nu \bar \nu
b \bar b$ in the $e^-_{\small L,R} e^+$ 
collisions could give important information about the interplay
of bremsstrahlung and fusion Higgs boson diagrams in the complete tree
level amplitude and could be employed as an essential Higgs identification test.
 
In the case of many irreducible background diagrams contributing to
the observables (total cross sections and distributions of the $b$-quarks)
it is important to separate reliably the signal and exclude the
situation when bias contribution could be misidentified as anomalous
signal. In the case under consideration the irreducible background can
be removed by the $M(b \bar b)$ cut around the Higgs peak. We observe an
interesting feature of bremsstrahlung-signal interference, which changes
sign as the energy increases and at some energy above the $M_Z+M_H$
threshold has ``interference zero''. At this point the experimentally
measurable value of polarized
asymmetry is defined only by individual contributions of two Higgs boson
'signal' diagrams and is directly related to the ratio of $ZZH$ and
$WWH$ couplings. One can expect about 2\% accuracy of the measurement for the 
polarization asymmetry at future linear colliders with reasonable integrated
luminosity.
 
\vspace{5mm}
\begin{center}
{\bf Acknowledgements}
\end{center}
 
The work was partially supported by 
INTAS grant 93-1180, RFBR grant 96-02-19773a
and Mombusho grant for International Scientific
Research Program No.07044097.

 
\newpage
\unitlength=0.8pt
\begin{figure}
\input{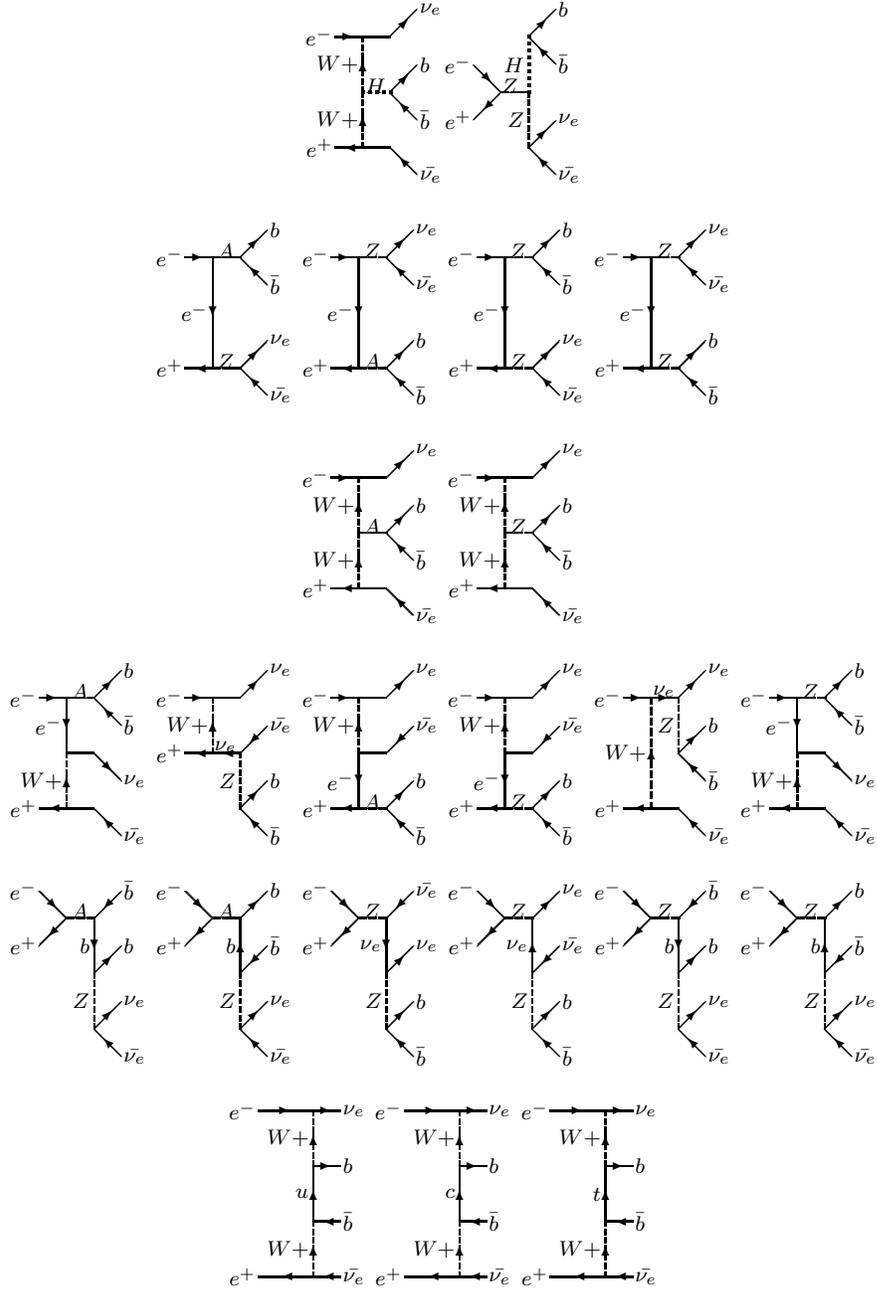}
\caption{Complete set of tree diagrams for the process
         $e^+ e^- \rightarrow \nu_e \bar \nu_e b \bar b$.}
\end{figure}
\unitlength=1pt
\unitlength=10mm
\begin{figure}[tbp]
\begin{picture}(11,13) 
\put(-4,0){\epsfbox{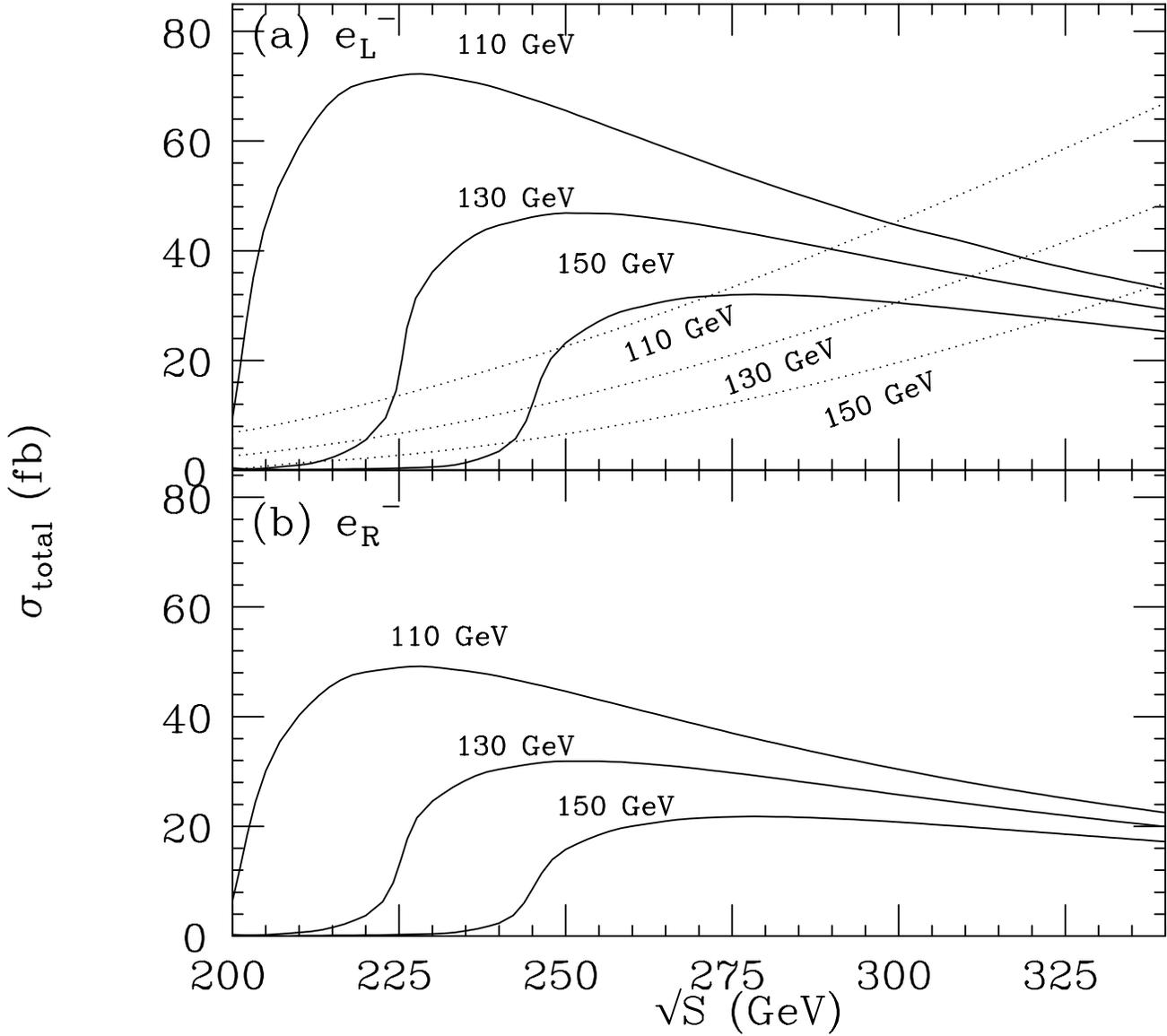}}
\end{picture}
\caption{{\bf (a)} Total cross sections calculated separately
for the bremsstrahlung (solid lines)
and fusion (dotted lines) mechanisms of Higgs boson production
in the process $e_L^- e^+ \rightarrow \nu \bar \nu b \bar b$,
$\nu=\nu_e, \nu_{\mu}, \nu_{\tau}$. Masses
of the Higgs boson $M_H=$ 110, 120 and 130 GeV, $m_b=$ 5 GeV, $M_Z=$ 91.16
GeV, $\Gamma_Z=$ 2.53 GeV, $\alpha=$ 1/128. 
{\bf (b)} Total cross sections for the
bremsstrahlung mechanism at the same parameter values for the process
$e^-_R e^+ \rightarrow \nu \bar \nu b \bar b$.}   
\end{figure}
\newpage
\begin{figure}[tbp]
\begin{picture}(10,12) 
\put(-4,0){\epsfbox{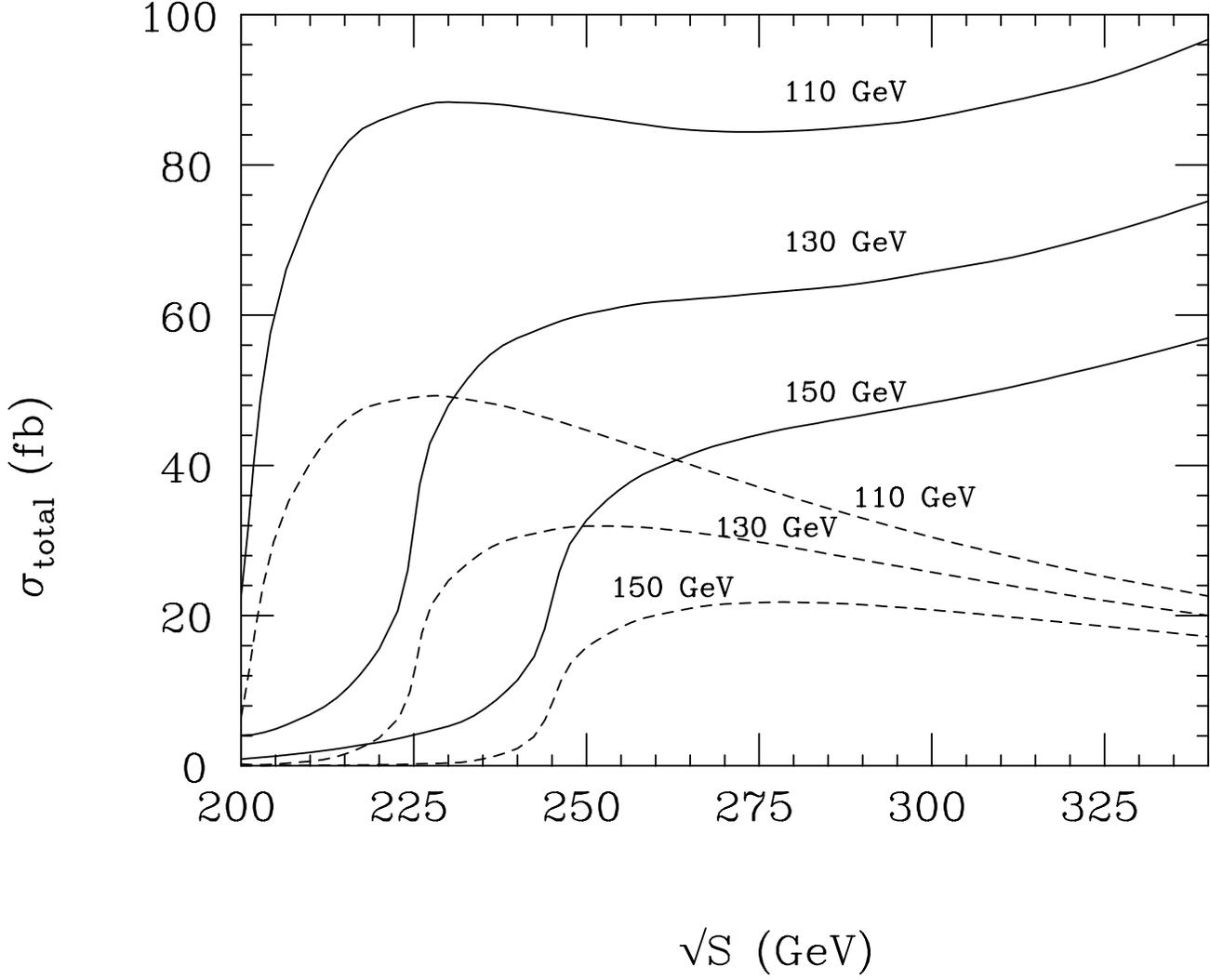}}
\end{picture}
\caption{Complete tree level cross sections vs energy for the processes
$e^-_L e^+ \rightarrow \nu \bar \nu b \bar b$ (solid lines) and
$e^-_R e^+ \rightarrow \nu \bar \nu b \bar b$ (dashed lines) for the
masses of Higgs boson $M_H=$ 110, 120 and 130 GeV. $M(b \bar b)$ cut of
$\pm1$ GeV around {$\protect M(b \bar b)=M_H$ is imposed}.}
\end{figure}
\newpage
\begin{figure}[tbp]
\begin{picture}(10,12)
\put(-4,0){\epsfbox{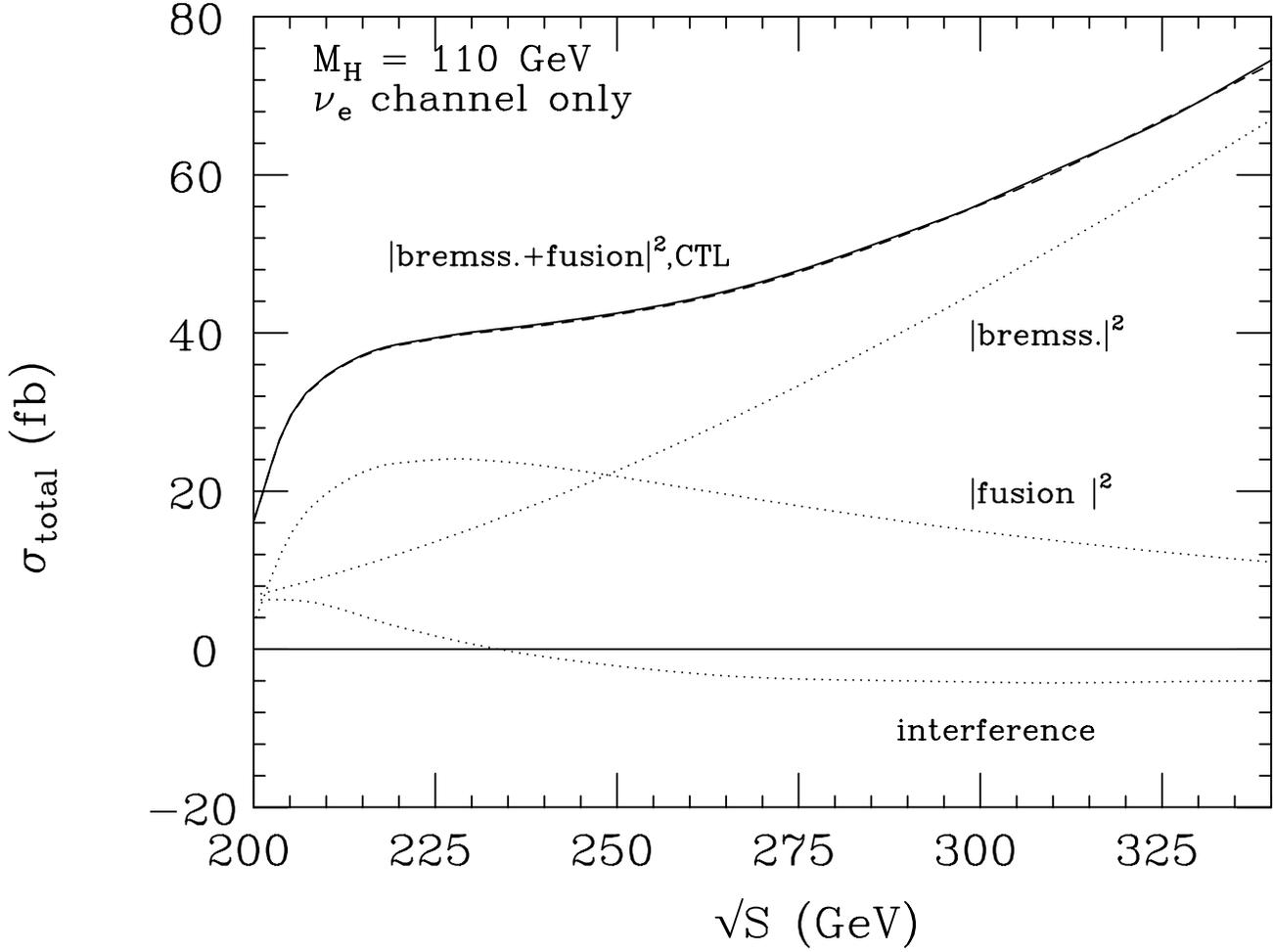}}
\end{picture}
\caption{The contributions of bremsstrahlung, fusion and bremsstrahlung-
fusion interference to the signal cross section 
for $M_H=$ 110 GeV, $\Gamma_H=$ 6.12 MeV are shown by dotted lines. 
The results of complete tree level calculation (CTL) and two signal 
diagrams squared sum are also shown by solid and dashed lines,
respectively.
The
(signal) interference term changes sign and is practically negligible around
{$\protect \sqrt{s} \sim $ 235 GeV}. For $M_H=$ 130  GeV, $\Gamma_H=$
7.26 MeV the 'interference zero' takes place at {$\protect \sqrt{s} \sim
$ 250 GeV}.}
\end{figure}
\newpage
\begin{figure}[tbp]
\begin{picture}(10,12)
\put(-4,0){\epsfbox{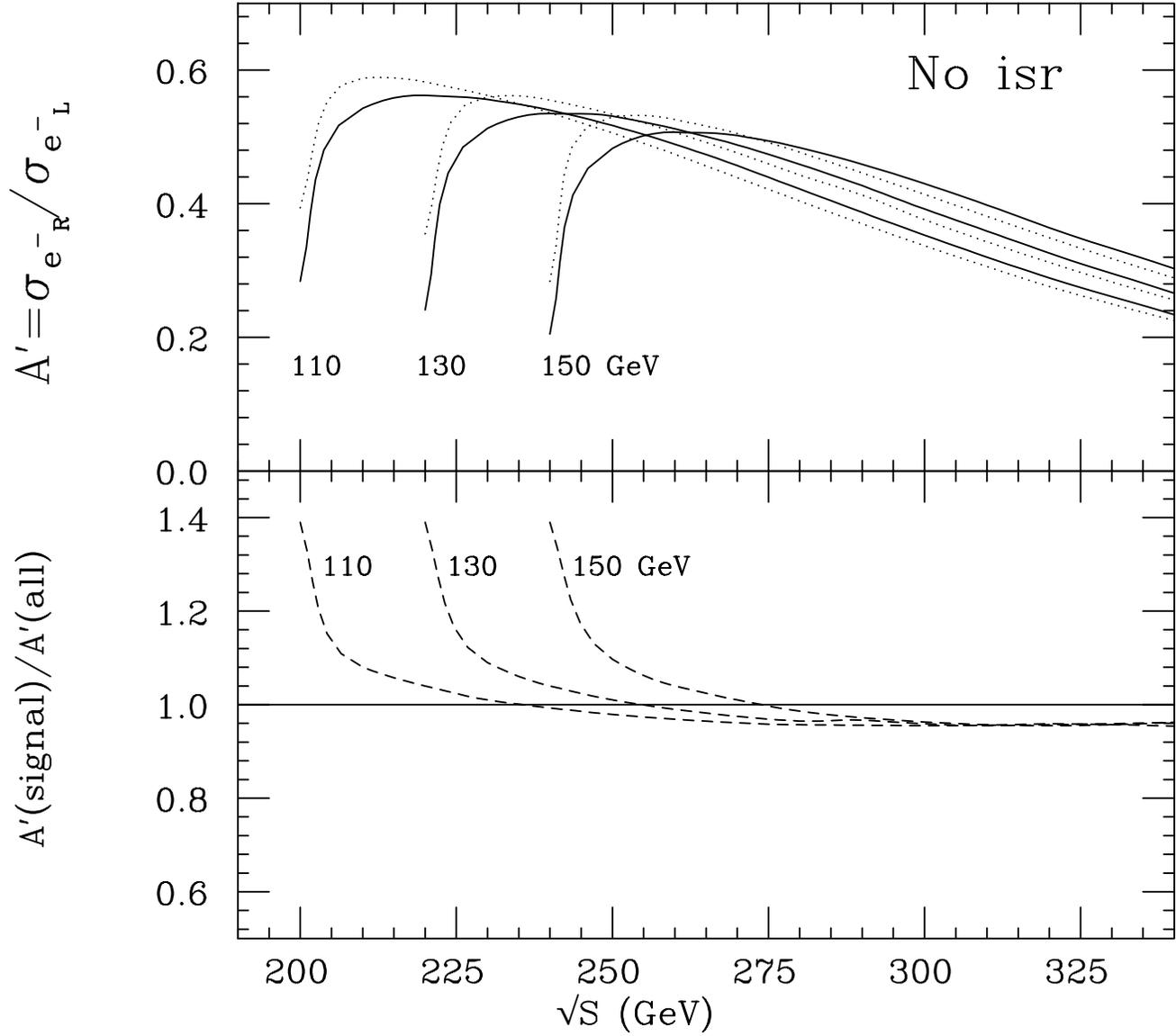}}
\end{picture}
\caption{The polarization asymmetry (upper plot, see formula (5)) 
for monochromatic beams,
$M_H=$ 110,120 and 130 GeV. $A'_{LR}$ calculated using signal diagrams only
(no signal interference)
is shown by a dotted line. The ratio of non-interfering signal diagrams  
to complete tree level cross section is shown in the lower plot. }
\end{figure}
\newpage
\begin{figure}[tbp]
\begin{picture}(10,12) 
\put(-4,0){\epsfbox{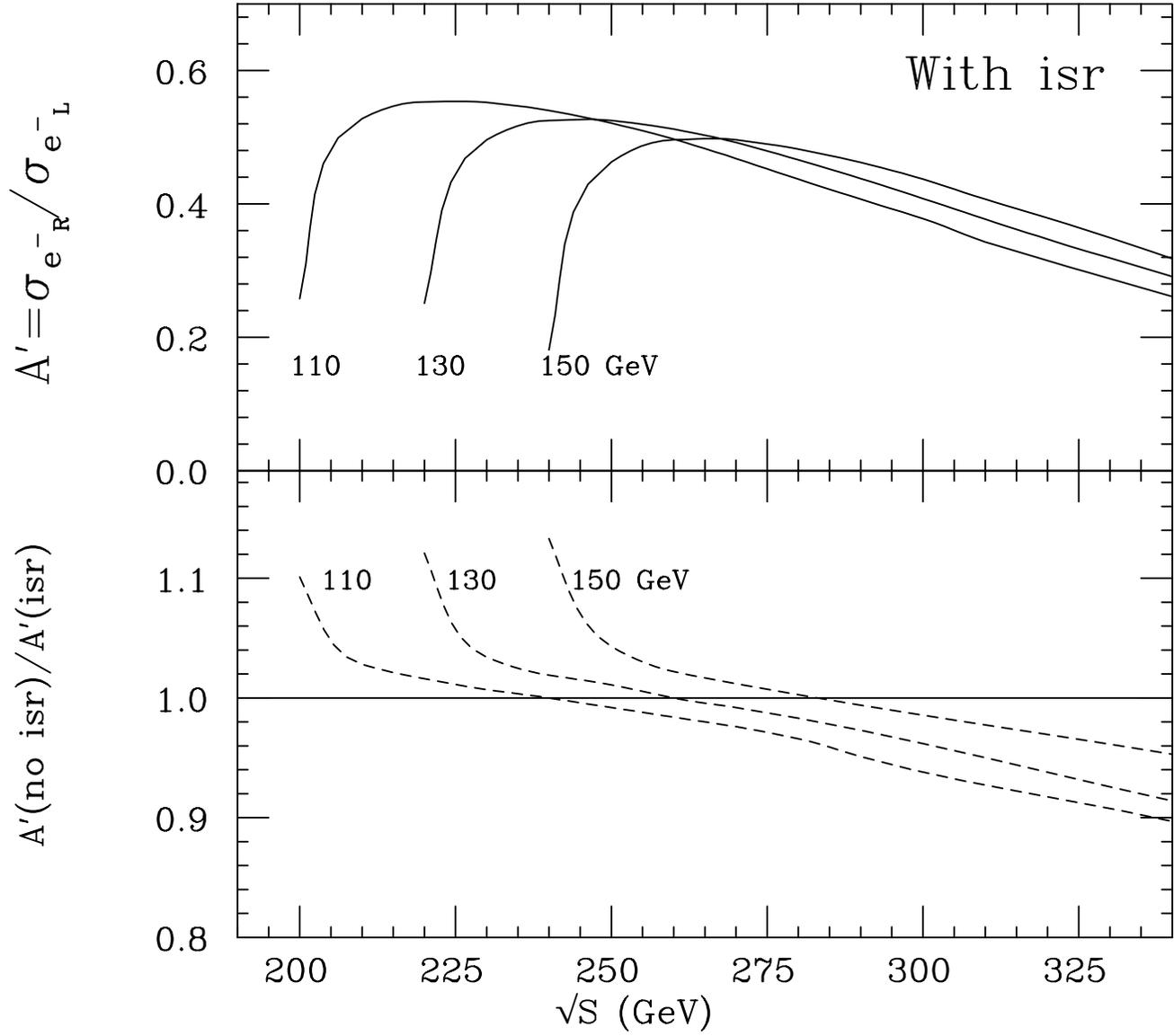}}
\end{picture}
\caption{The polarization asymmetry (see formula (5)) calculated with
initial state radiative corrections, $M_H=$ 110, 120 and 130 GeV.} 
\end{figure}
\newpage
\begin{figure}[tbp]
\begin{picture}(10,12) 
\put(-4,0){\epsfbox{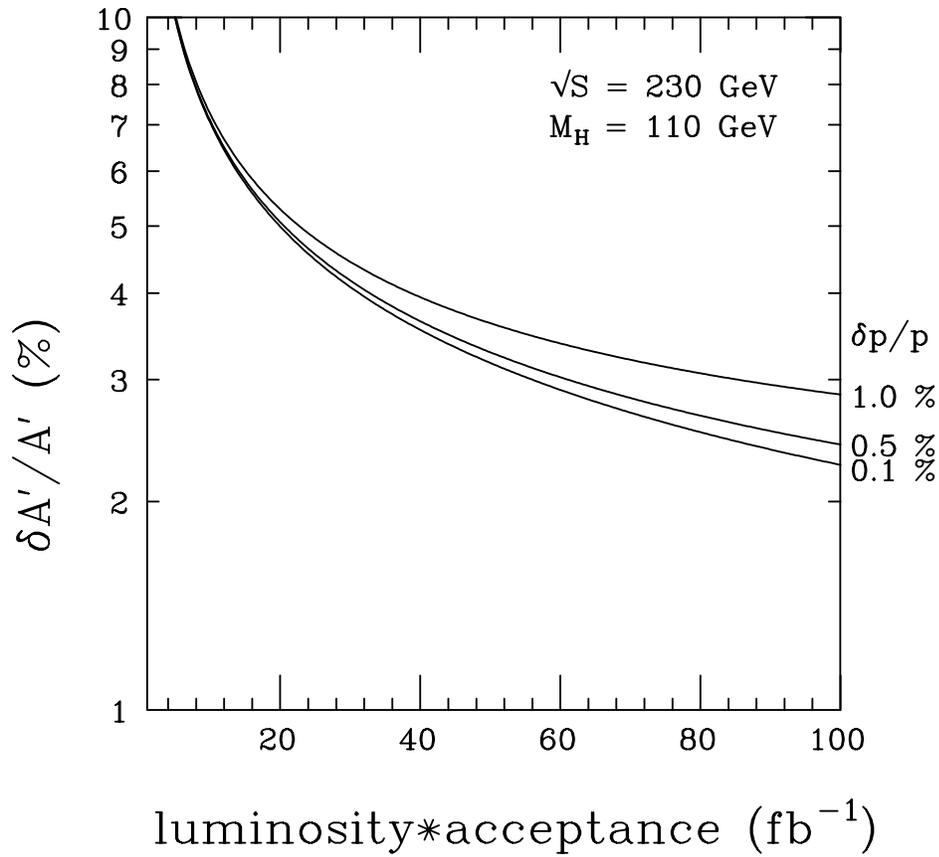}}
\end{picture}
\caption{The expected accuracy of the $A'_{LR}$ measurement at 
CM energy of 230 GeV and $M_H$=110 GeV.
We assume 0.1\% accuracy for the calculation of functions $F$ 
in equation (5) and 0.2\% accurracy for $\cos^2 \vartheta_W$ measurement.
Three different values of the polarization measurement accuracy were used
in the simulation.}
\end{figure}
\end{document}